\documentclass[prd,nofootinbib,preprint,eqsecnum,12pt]{revtex4-2}
\usepackage{amssymb,amsmath}
\usepackage{amsfonts,mathrsfs}
\usepackage{epstopdf}
\usepackage{graphicx}
\newcommand{\be}{\begin{equation}}
\newcommand{\ee}{\end{equation}}
\newcommand{\bea}{\begin{eqnarray}}
\newcommand{\eea}{\end{eqnarray}}
\newcommand{\bes}{\begin{subequations}}
\newcommand{\ees}{\end{subequations}}
\newcommand{\la}{\langle}
\newcommand{\ra}{\rangle}
\newcommand{\w}{\omega}
\def\kb{\kappa_b}
\def\kc{\kappa_c}
\def\kn{\kappa_N}
\begin{document}
\author{Paul~R.~Anderson}
\email{anderson@wfu.edu}

\author{Shohreh Gholizadeh Siahmazgi}
\email{ghols18@wfu.edu}

\author{Zachary P.\ Scofield}

\affiliation{Department of Physics, Wake Forest University, Winston-Salem, North Carolina 27109, USA}

\title{\Large \bf   Infrared Effects and the Unruh State}

\begin{abstract}

Detailed behaviors of the modes of quantized scalar fields in the Unruh state for various eternal black holes in two dimensions are investigated.  It is shown that the late-time behaviors of some of the modes of the quantum fields and of the symmetric two-point function are determined by infrared effects.  The nature of these effects depends upon whether there is an effective potential in the mode equation and what form this potential takes.  Here, three cases are considered, one with no potential and two with potentials that are nonnegative everywhere and are zero on the event horizon of the black hole and zero at either infinity or the cosmological horizon.  Specifically, the potentials are a delta function potential and the potential that occurs for a massive scalar field in Schwarzschild-de Sitter spacetime. In both cases, scattering effects remove infrared divergences in the mode functions that would otherwise arise from the normalization process.  When such infrared divergences are removed, it is
found that the modes that are positive frequency with respect to the Kruskal time on the past black hole horizon
approach zero in the limit that the radial coordinate is fixed and the time coordinate goes to infinity.  In contrast, when there is no potential and thus infrared divergences occur, the same modes approach nonzero constant values in the late-time limit when the radial coordinate is held fixed.
The behavior of the symmetric two-point function when the field is in the Unruh state is investigated for the case of a delta function potential in certain asymptotically flat black hole spacetimes in two dimensions.  The removal of the infrared divergences in the mode functions results in the elimination of terms that grow linearly in time.

  \end{abstract}

\maketitle

\section{Introduction}

In his original calculation of black hole evaporation~\cite{hawking:1975}, Hawking considered the case of a black hole that forms from the collapse of matter in an asymptotically flat spacetime.  He found that the state of a quantum field that corresponds to the vacuum state at early times is not a vacuum state at late times, but instead consists, at large distances from the black hole, of a flux of particles in a thermal distribution.  Particle production of this type requires a time-dependent geometry and thus does not occur in static spacetimes such as the Schwarzschild solution to Einstein's equations.  However, it is mathematically much simpler to study quantum effects in static and stationary spacetimes, in part because the mode equations for free quantum fields are separable.

There are typically three states for the quantum fields that are considered for eternal black holes.  These are the Boulware state~\cite{boulware}, the Hartle-Hawking state~\cite{hartle-hawking}, and the Unruh state~\cite{unruh:1976}.  A state for a free quantum field can be specified by choosing a complete set of solutions to the mode equation for the quantum field.  Then the field can be expanded in terms of these with creation and annihilation operators as the coefficients.  See e.g. Eq.~\eqref{phi-first} below.

The Boulware state can be specified by choosing a complete set of modes that are positive frequency with respect to the usual time coordinate $t$.  Specifically, at past null infinity (or the past cosmological horizon if there is one), some of these modes are initially in-going modes.  The rest are out-going modes on the past black hole horizon.  At past and future null infinity in a spacetime  that is asymptotically flat, the Boulware state corresponds to a true vacuum state in the sense that there are no particles that a particle detector on an inertial trajectory would detect.  However, there is a serious problem in choosing it for the
state of the quantum field because the stress-energy tensor for the field diverges on the past
and future horizons of the black hole.

The Hartle-Hawking state~\cite{hartle-hawking} can be defined in terms of the union of sets of positive frequency modes with respect to the usual Kruskal coordinates on the past and future horizons of the black hole~\cite{israel}.  The resulting two-point functions are those for a thermal state at the black hole temperature.
For an asymptotically flat black hole spacetime, the Hartle-Hawking state results in a stress-energy tensor for the quantum field that is regular
on both the past and future horizons.  However, it is not compatible with an asymptotically flat spacetime because the stress-energy tensor approaches that for a thermal distribution of particles in flat space far from the black hole.  Instead, it is best thought of as the state that would describe what happens if one placed a perfectly reflecting, spherically symmetric mirror around a spherically symmetric black hole.

If there is a cosmological horizon, then one can also define a corresponding Hartle-Hawking state using Kruskal coordinates appropriate for the cosmological horizon.  However, the two Hartle-Hawking states are not compatible because the temperatures associated with the horizons are different.  As a result, the stress-energy tensor for the quantum field will diverge either on the cosmological horizon if the state is defined with respect to the black hole horizon Kruskal coordinates, or it will diverge on the black hole horizon if the state is defined with respect to the cosmological horizon Kruskal coordinates.

For an asymptotically flat, static black hole spacetime, the Unruh state consists of the union of positive frequency modes with respect to the Kruskal coordinates on the past black hole horizon and the modes that are positive frequency with respect to the usual time coordinate $t$ on past null infinity. For static black holes, the Unruh state~\cite{unruh:1976} is arguably the most important one for studying the Hawking effect because the Unruh state was designed to be a state which mimics the late-time behavior of quantized fields  that was predicted by Hawking~\cite{hawking:1975} for a black hole that forms from collapse.  In particular it has the same flux of particles at future null infinity with the same spectrum.  The stress-energy tensor for the quantum field is finite on the future black hole horizon, but diverges on the past horizon.  This is generally not thought to be a problem, because only the future horizon exists for a black hole that forms from collapse.
  The Unruh state has also been defined for Schwarzschild-de Sitter, SdS, spacetime~\cite{Markovic:1991} (see also~\cite{Tadaki:1990a,Tadaki:1990b}) which has both a black hole and a cosmological horizon. The modes at past null infinity are replaced by modes that  are positive frequency on the past cosmological horizon with respect to the Kruskal coordinates appropriate for that horizon. This has an obvious generalization to any case in which there is both a black hole and a cosmological horizon.

In this paper we consider the Unruh state for static black hole metrics in two dimensions (2D).  We focus on the late-time behaviors of the mode functions that are positive frequency with respect to the Kruskal coordinates on the past black hole horizon.  For simplicity we call these the Kruskal modes.    We find that these behaviors are strongly influenced by the infrared behaviors of the Boulware modes that are outgoing modes on the past black hole horizon.  In addition, we compute the symmetric two-point function, also called the Hadamard Green's function, and find that its late-time behavior when the points are split in the space direction is strongly influenced by the same infrared behaviors of the Boulware modes.

As is well known, the mode equation for a scalar field in a static black hole spacetime such as Schwarzschild spacetime in four dimensions (4D) has an effective potential that depends on the radial coordinate $r$.  This effective potential results in the scattering of the mode functions.  In its absence, the general solution is made up of an arbitrary function of left moving waves and an arbitrary function of right moving waves.  If the wave equation is separable in the usual $t$ and $r$ coordinates, one can use the solutions to define the Boulware state~\cite{boulware} which is a true vacuum state at past and future null infinity.

If there is no scattering, it is easy to show that a Kruskal mode approaches a nonzero constant at late times if the radial coordinate is held fixed.  If there is scattering due to an effective potential, then it is nontrivial to find the behavior of a Kruskal mode away from the past black hole horizon.  There are different approaches that can be taken.  Here we expand the Kruskal modes in terms of the Boulware modes that are positive frequency on the past black hole horizon.  We find, for the cases considered, that the infrared divergences in the Boulware modes are removed by scattering effects, and that when these are used to compute the Kruskal modes, the Kruskal modes vanish in the late-time limit if the radial coordinate is held fixed.

The specific cases we consider are the massless minimally coupled scalar field where there is no scattering and the massless minimally coupled scalar field when there is a delta function potential, both in a spacetime with a general 2D static black hole metric, and the massive minimally coupled scalar field in SdS with a 2D metric. The delta function potential has a positive coefficient and can be thought of as an extreme approximation for the potential for the massive scalar field in SdS  with a 2D metric as well as the massless minimally coupled scalar field in 4D static black hole spacetimes that are asymptotically flat.

For a massless minimally coupled scalar field
 in a spacetime with a 2D general static black hole metric, it was shown in~\cite{a-t} that the symmetric two-point function grows linearly in time when the points are split in the radial direction for a time coordinate that is regular on the future horizon and is regular on future null infinity for an asymptotically flat spacetime or the future cosmological horizon for SdS.  In~\cite{unruh:1976}, two different formulations of the Unruh state were given along with the statement that the vacuum states for the two formulations are equivalent (see also~\cite{fulling:1977}).  One way to verify this statement is to use a
 Bogolubov transformation to expand the Kruskal modes in terms of the Boulware ones and then use the result to compute the two-point function.  If the order of integration in the resulting triple integral is changed in the right way, the resulting integrand is equivalent to that obtained using the second formulation.\footnote{The evaluation of the triple integral in this way was done previously in~\cite{bec}, however, no mention was made of the two formulations of the Unruh state.  It was also done in the first two arXiv versions of~\cite{a-t}, but does not appear in the last two arXiv versions or the published version of that paper.}    In the case of a massless minimally coupled scalar field in 2D, there is an infrared divergence in the Boulware modes that probably prevents one from changing the order of integration.  However, for the delta function potential,  the mode functions in the Boulware state have no infrared divergences and there is no problem with changing the order of integration.

 In Sec. II, the solutions to the mode equation are discussed for the cases considered.  In Sec. III, the computation of the symmetric two-point function when there is no scattering is reviewed for eternal black hole spacetimes with 2D metrics.  The connection mentioned above between the two formulations of the Unruh state is shown.  For a delta function potential, the two-point function is computed explicitly in both formulations for several specific cases and it is shown that they are equivalent.  Sec. IV contains a summary of our results.  Throughout we use units such that $\hbar = G = c = 1$.

\section{Mode functions}
\label{sec:modes}

For the usual static black hole solutions to Einstein's equations in 4D, the metric can be written in the form:
\be ds^2 =  -f(r) dt^2 + \frac{dr^2}{f(r)} + r^2 d \Omega^2  =  f(r) [-dt^2 + dr_*^2]   + r^2 d \Omega^2     \;,  \ee
with $d \Omega^2$ the angular part of the metric and
\be r_* \equiv \int^r \frac{dr'}{f(r')} \;. \ee
In general, a free scalar field can have a coupling to the scalar curvature of the form $\xi R$ for some dimensionless constant $\xi$.  In this paper, we restrict attention to the case of minimal coupling, $\xi = 0$.  We also restrict our attention to the 2D form of the metric which is obtained by setting $d \Omega = 0$.  In the above coordinates, the mode equation for a massive scalar field is then
\be \left[-\frac{\partial^2}{\partial t^2} + \frac{\partial^2 }{\partial r_*^2} - V(r_*) \right] \psi(t,r_*) = 0  \;, \ee
with $V(r_*) = m^2 f(r_*)$.\footnote{Note that in 4D Schwarzschild spacetime the potential takes the form $V(r) = \left(1-\frac{2M}{r}\right) \left[\frac{2M}{r^3} + \frac{ \ell (\ell+1)}{r^2} + m^2 \right] $.}
One important property of the potential is that it vanishes at the black hole horizon where $r_* \to - \infty$ and at infinity (or the cosmological horizon) where $r_* \to + \infty$.  For the cases we consider, the potential is nonnegative in between and has just one peak. This is also the property it has for the massless field in 4D for Schwarzschild and Reissner-Nordstrom spacetimes.
For the massless field in 2D, $V = 0$ and the general solution is
\be \psi = g(u) + h(v) \;, \ee
with $u = t - r_*$, $v = t + r_*$, and $g$ and $h$ arbitrary functions.

The mode functions can be normalized using the usual scalar product on the Cauchy surface that consists of the past black hole horizon and either
past null infinity or the past cosmological horizon.  Separation of variables results in modes of the form
\be h_\w = \frac{1}{\sqrt{4 \pi \w}} e^{-i \w t} \chi_\w(r_*) \label{modes-sep-var} \;, \ee
with
\be \frac{d^2 \chi_\w}{d r_*^2} + (\w^2 - V(r_*)) \chi_\w = 0 \;. \label{mode-eq-chi} \ee
The Boulware state consists of the union of the set of modes that on the past black hole horizon have the form
\be h^b_\w = \frac{1}{\sqrt{4 \pi \w}} e^{-i \w u} \;, \label{hb-def} \ee
and the set of modes that on past null infinity or the past cosmological horizon have the form
\be h^{(\mathscr{I}^{-}, \, c)}_\w = \frac{1}{\sqrt{4 \pi \w}} e^{-i \w v} \;. \label{hI-hc-def} \ee

One can define Kruskal coordinates in the usual way such that outside the past and future black hole horizons
\be U_b = - \frac{1}{\kappa_b} e^{- \kappa_b u} \;, \qquad  V_b = \frac{1}{\kappa_b} e^{\kappa_b v} \;. \label{UbVb} \ee
If there are past and future cosmological horizons, then inside of them
\be U_c =  \frac{1}{\kappa_c} e^{\kappa_c u} \;, \qquad  V_c = -\frac{1}{\kappa_c} e^{-\kappa_c v} \;, \label{UcVc} \ee
with $\kappa_b$ and $\kappa_c$ the surface gravities of the black hole and cosmological horizons respectively.

A set of modes exists whose form on the past black hole horizon is given by
\be p^b_\w = \frac{1}{\sqrt{4 \pi \w}} e^{-i \w U_b}\;, \label{pb} \ee
and a set exists whose form on the future black hole horizon is given by
\be q^b_\w =  \frac{1}{\sqrt{4 \pi \w}} e^{-i \w V_b}\;. \ee
If there is a cosmological horizon then there is a set of modes whose form on the past cosmological horizon is given by
\be p^c_\w = \frac{1}{\sqrt{4 \pi \w}} e^{-i \w V_c}\;, \ee
and there is a set of modes whose form on the future cosmological horizon is given by
\be q^c_\w =  \frac{1}{\sqrt{4 \pi \w}} e^{-i \w U_c}\;. \ee

The Unruh state for a black hole in an asymptotically flat spacetime is defined by the union of the modes $p^b_\w$ and $h^{\mathscr{I}^{-}}_\w$.  In a spacetime with a cosmological horizon it is the union of $p^b_\w$ and $p^c_\w$.  The Hartle-Hawking state in an asymptotically flat spacetime is defined
by the union of $p^b_\w$ and $q^b_\w$.  There are two possible Hartle-Hawking states in a spacetime with a cosmological horizon, however neither
is well behaved at both future horizons because the temperatures of those horizons are different.

When the potential is zero, the above sets of modes take the above forms everywhere.  When it is nonzero, then scattering occurs and the radial mode equation has nontrivial solutions.  Since the mode equation in Kruskal coordinates cannot be solved using separation of variables, it becomes difficult to find the form of $p_\w$ and $q_\w$ away from the horizons where they are defined.  One way to do this is to expand these modes in terms of the Boulware modes $h_\w$ which can be found using separation of variables.  For example,
\be p^b_\w = \int_0^\infty  d \w \;[\alpha^b_{\w \w'} h^b_{\w'} + \beta^b_{\w \w'} h^{b *}_{\w'} ] \;. \label{pbeq} \ee
Using the usual scalar product and the orthonormality of the modes $h^b_\w$ with respect to that scalar product, one finds\footnote{These coefficients have been previously derived in~\cite{bec},~\cite{null-shell-method}, and~\cite{null-shell-proceedings}.  However, there are mistakes and misprints in~\cite{bec} and~\cite{null-shell-method}.}
\bes \bea \alpha^b_{\w \w'} &=& -i \int_{-\infty}^\infty du \; [ p^b_\w \partial_u h^{b *}_{\w'} - (\partial_u p^b_\w) h^{b *}_{\w'} ]  \nonumber \\
                      &=& \frac{1}{2 \pi \kappa_b} \sqrt{\frac{\w'}{\w}} \, \frac{\Gamma(\delta - i \frac{\w'}{\kappa_b})}{ \left(-i \frac{\w}{\kappa_b}+ \epsilon \right)^{-i \frac{ \w'}{\kappa_b}}}  \;, \\
  \beta^b_{\w \w'} &=& i \int_{-\infty}^\infty du \; [ p^b_\w \partial_u h^{b }_{\w'} - (\partial_u p^b_\w) h^{b }_{\w'} ]  \nonumber \\
                     &=&  \frac{1}{2 \pi \kappa_b} \sqrt{\frac{\w'}{\w}} \, \frac{\Gamma( \delta + i \frac{\w' }{\kappa_b})}{ \left(-i \frac{\w}{\kappa_b} + \epsilon \right)^{i \frac{\w'}{\kappa_b}}} \;.
\eea  \label{alpha-beta-b} \ees
Here $\delta$ and $\epsilon$ are small positive integrating factors that should be set to zero at the end of a calculation.
If a cosmological horizon is present, the Bogolubov coefficients for $p^c_\w$ can be obtained by substituting $\kappa_c$ for $\kappa_b$ everywhere in~\eqref{alpha-beta-b}.

\subsection{No Potential}
\label{sec:nopot}

If the potential is zero, then the Kruskal mode functions for the Unruh state take the form that they have on the past horizon everywhere.
This allows us to easily determine the late-time behaviors of these modes and also to examine in detail the expansion~\eqref{pbeq}.  We shall focus
on $p^b_\w$.  However, the results immediately generalize to $q^b_\w$ as well as $p^c_\w$ and $q^c_\w$.

First, note that $u \to \infty$ on the future black hole horizon and also at future timelike infinity for fixed $r_*$.  From~\eqref{UbVb} it is clear that $U \to 0$ in this limit.  Therefore $p^b_\w \to (4 \pi \w)^{-1/2}$.
In contrast, $u \to -\infty$ on past null infinity (or the past cosmological horizon) and $U \to -\infty$ as well in this limit.  Then $p^b_{\w}$ oscillates rapidly as the limit is approached and this causes a typical wave packet made of these modes to vanish in the limit $U \to - \infty$.

It is interesting to examine~\eqref{pbeq} in this case using~\eqref{alpha-beta-b}.  Since $h^b_{\w'} = \frac{e^{-i \w' u}}{\sqrt{4 \pi \w'}}$ everywhere, one can easily see that for $\w' < 0$,
\be [\beta^b_{\w \w'} h^{b *}_{\w'}]_{\w' \to - \w'}  = \alpha^b_{\w \w'} h^b_{\w'} \;, \label{beta-condition} \ee
where the quantities on the right hand side are to be evaluated at $\w' < 0$.  Then,
\be p^b_\w = \int_{-\infty}^\infty d \w' \; \alpha_{\w \w'}^b h^b_{\w'} =  \frac{1}{2 \pi \kappa_b \sqrt{4 \pi \w}} \int_{-\infty}^\infty d \w' \, \frac{\Gamma(\delta - i \frac{\w'}{\kappa_b})}{ \left(-i \frac{\w}{\kappa_b}+ \epsilon \right)^{-i \frac{ \w'}{\kappa_b}}} e^{-i \w' u} \label{pb-one-integral}  \;. \ee
The integral can be computed using complex integration.  The poles of the gamma function are at
\bea \w' &=& -i \kappa_b \delta \;, \nonumber \\
     \w' &=& - i \kappa_b n \;, \qquad n = 1, 2, \ldots
\eea
It is straight-forward to show that the sum of the residues gives~\eqref{pb} for all values of $u$ as expected.

One can also examine the integrand evaluated along the real axis. For large $\w'$ one can use Sterling's approximation to obtain
\bes \bea p^b_\w &=& \frac{e^{i \pi/4}}{\sqrt{2 \pi \kappa_b} \sqrt{4 \pi \w}} \int_0^\infty \frac{d \w'}{\sqrt{\w'}} \, e^{i \Omega_A}
           +  \frac{e^{-i \pi/4}}{\sqrt{2 \pi \kappa_b} \sqrt{4 \pi \w}} \int_0^\infty \frac{d \w'}{\sqrt{\w'}} \, e^{-\frac{\pi \w'}{\kappa_b}}  e^{i \Omega_B} \;, \\
           \Omega_A &=& \w'\left(\frac{1}{\kappa_b} - u \right) - \frac{\w'}{\kappa_b} \log\frac{\w'}{\w}  \;, \\
           \Omega_B &=& \w'\left(-\frac{1}{\kappa_b} + u \right) + \frac{\w'}{\kappa_b} \log\frac{\w'}{\w} \;.
    \eea \ees
Both integrals have a stationary phase point at \be \w'_s = \w e^{-\kappa_b u} = -\w  \kappa_b U_b \;.  \ee The stationary phase approximation gives to leading order \be (p^b_\w)_s = \frac{1}{\sqrt{4 \pi \w}} \left[ e^{-i \w U_b} + e^{i \w U_b} e^{\pi \w U} \right] \label{pb-s} \;. \ee

To analyze this result, first note that
Sterling's approximation is only valid at the stationary phase point if $\w'_s \gg \kappa_b$ so that $\w e^{-\kappa_b u} = - \w \kappa_b U_b \gg \kappa_b$.
Therefore, the second term is always negligible when~\eqref{pb-s} is valid.  The first term is just the exact expression for $p^b_\w$ that
is valid everywhere when there is no scattering.
Second, note that for fixed $\w$ the approximation, and therefore~\eqref{pb-s}, is only valid at early times.  In contrast for fixed $u$, the approximation is always valid for large enough values of $\w$, but the lower limit cutoff for validity gets larger as $u$ increases.

For $\w e^{-\kappa_b u} \lesssim \kappa_b$, the dominant contribution to the integral in~\eqref{pb-one-integral} comes from the region near $\w' = 0$.  This region gets
smaller as $u$ increases for fixed $\w$.  However, there is a singularity at $\w' = 0$ coming from $\Gamma\left(\delta - i \frac{\w'}{\kappa_b} \right)$. As a result, integrating over this region gives a constant contribution that is equal to $\frac{1}{\sqrt{4 \pi \w}}$.  This is exactly the value of $p^b_\w$ in the limit $u \to \infty$ or $U \to 0$.

Thus we see that it is the infrared behavior of the wave packet of Boulware modes that determines the late-time behavior of $p^b_\w$.  Since the form of the Bogolubov coefficients is exactly the same for the other Kruskal modes, similar results occur for them when there is no scattering.   As is shown below, the different infrared behavior of this integrand when the infrared divergences in the Boulware modes are removed results in a completely different late-time behavior for $p^b_\w$.

\subsection{Scattering due to a Potential}

In general if there is a potential then scattering effects will occur.  The general form of the modes that can be obtained using separation of variables is given in~\eqref{modes-sep-var}, and from this equation it is clear that the scattering effects come from solutions to the radial mode equation~\eqref{mode-eq-chi}.  In this paper we assume that the potential vanishes at both the past black hole horizon and either past null infinity or the past cosmological horizon.  Then~\eqref{hb-def} gives the behavior of one set of the Boulware modes on the past black hole horizon.  Similarly,~\eqref{hI-hc-def} gives the behavior of the other set on either past null infinity or the past cosmological horizon.  Rather than working with these modes directly, it is useful to consider  two linearly independent solutions to the radial mode equation that in the limit $r_* \to \infty$ have the form
\be \chi^\infty_R \to e^{i \w r_*} \;, \qquad  \chi^\infty_L \to e^{-i \w r_*}  \;. \label{chi-R-chi-L-infinity} \ee
Because of scattering effects, in the limit $r_* \to -\infty$ these solutions have the general form
\bes \bea \chi^\infty_R &\to& E_R e^{i \w r_*} + F_R e^{-i \w r_*}  \;, \\
    \chi^\infty_L &\to& E_L e^{i \w r_*} +  F_L e^{-i \w r_*} \;, \eea \label{chi-R-L-minfinity} \ees
\noindent where the coefficients of the exponentials are complex functions of $\w$.

The combinations of $\chi^\infty_R$ and $\chi^\infty_L$ that give the radial parts of the Boulware modes $\chi^b_\w$ and either $\chi^c_\w$ or $\chi^{\mathscr{I}^{-}}_\w$
are
\bes \bea \chi^b_\w &=& \frac{\chi^\infty_R}{E_R} \;, \label{chi-b-scatt} \\
           \chi^{(\mathscr{I}^{-},\, c)}_\w &=& \chi^\infty_L - \frac{E_L}{E_R} \chi^\infty_R  \;. \label{chi-I-c-scatt}  \eea \label{chi-b-I-c-scatt} \ees

\subsection{Delta Function Potential}
\label{sec:delta}

Another case for which the radial mode equation can be solved analytically is a massless minimally coupled scalar field   for a static black hole with a 2D metric when the potential is
\be V(r_*) = \lambda \delta(r_*)  \;. \label{delta-potential} \ee
Taking $\lambda > 0$ gives a potential in the same form as that for a massive field in SdS spacetime with a 2D metric and a massless minimally coupled scalar field in 4D Schwarzschild spacetime.  That is, the potential is zero on the black hole horizon and either the cosmological horizon or infinity, and it is nonnegative.

Solving~\eqref{mode-eq-chi} with~\eqref{delta-potential}, we find for $r_* > 0$ that
 \be \chi^\infty_R = e^{i \w r_*} \;, \qquad \chi^\infty_L = e^{-i \w r_*} \;, \label{delta-r-greater}  \ee
while for $r_* < 0$
\bea \chi^\infty_R &=& E_R e^{i \w r_*} + F_R e^{- i \w r_*} \;, \nonumber \\
\chi^\infty_L &=& E_L e^{i \w r_*} + F_L e^{- i \w r_*} \;, \label{delta-r-lesser} \eea
with
\bea  E_R &=& 1 + i \frac{\lambda}{2 \w} \;, \nonumber \\
      F_R &=& -i \frac{\lambda}{2 \w} \;, \nonumber \\
      E_L &=& i \frac{\lambda}{2 \w} \;, \nonumber  \\
      F_L &=& 1 - i \frac{\lambda}{2 \w} \;.  \label{delta-scatt-coeff}
\eea

Substituting these results into~\eqref{chi-b-I-c-scatt} one finds that the mode functions for the Boulware state are
\bes \bea \sqrt{4 \pi \w}\; h^b_\w  &=&  \theta(-r_*) \left[ e^{-i \w u} - \frac{\frac{i \lambda}{2}}{\left(\w + \frac{i \lambda}{2} \right)} e^{-i \w v} \right]
  + \theta(r_*) \frac{\w}{\left(\w + \frac{i \lambda}{2} \right)} e^{-i \w u} \;, \label{h-b-delta} \\
  \sqrt{4 \pi \w} \; h^{(\mathscr{I}^{-},c)}_{\w} &=& \theta(-r_*) \frac{\w}{\left(\w + \frac{i \lambda}{2} \right)} e^{-i \w v}  + \theta(r_*) \left[ e^{-i \w v} - \frac{\frac{i \lambda}{2}}{\left(\w + \frac{i \lambda}{2} \right)} e^{-i \w u} \right] \;. \label{h-I-c-delta} \eea \label{h-b-I-c-delta} \ees
Note that scattering effects have removed the infrared divergences in these modes.

To compute the modes that are positive frequency with respect to the Kruskal time on the past horizon, we first note that~\eqref{beta-condition} is satisfied by~\eqref{h-b-delta}.  Then substituting~\eqref{h-b-delta} into the integral after the first equals sign in~\eqref{pb-one-integral}, one finds
\bea p^b_\w  & =& \int_{-\infty}^{\infty} d \w'
\alpha^b_{\w \w'} h^b_{\w'} \nonumber \\
 &=& \; \frac{1}{4 \pi^{3/2} \kappa_b \sqrt{\w}} \int_{-\infty}^{\infty}d\w'\;\Gamma\left(\delta-i\frac{\w'}{\kappa_b}\right)\left(-i\frac{\w}{\kappa_b}+\epsilon\right)^{i \frac{\w'}{\kappa_b}}\Big\{\theta (-r_*)\big(e^{-i\w' u}-\frac{\frac{i\lambda}{2}}{\w'+\frac{i\lambda}{2}}e^{-i\w'v}\big)\nonumber \\ &&
+\theta (r_*)\frac{\w'e^{-i\w' u}}{\w'+\frac{i\lambda}{2}}\Big\}  \;.  \label{pb-delta} \eea
This expression has the same poles on the negative imaginary axis as the second integral in~\eqref{pb-one-integral}.  However, it also has one more singularity at $\w' = - i \frac{\lambda}{2}$.  Evaluating the integral using complex integration and setting $\epsilon$ to zero gives
\bea p^b_\w  & =& \Big\{-\frac{\lambda}{4 \kappa_b \sqrt{\pi \w}}\Gamma\left(-\frac{\lambda}{2 \kappa_b}\right)\left(-i\frac{\w}{\kappa_b}\right)^{\frac{\lambda}{2 \kappa_b}}e^{-\frac{\lambda v}{2}}\nonumber \\ &&
+\frac{1}{\sqrt{4\pi\w}}\sum_{n=0}^{\infty} \frac{(-1)^n}{n!}\left(-i\frac{\w}{\kappa_b}\right)^n \Big(e^{-\kappa_b u n} +\frac{\frac{\lambda}{2 \kappa_b}}{ n -\frac{\lambda}{2\kappa_b}}e^{-\kappa_b v n}\Big)\Big\}\; \theta(-r_*)\nonumber \\ &&
+\Big\{-\frac{\lambda}{4 \kappa_b \sqrt{\pi \w}}\Gamma\left(-\frac{\lambda}{2 \kappa_b} \right)\left(-i\frac{\w}{\kappa_b}\right)^{\frac{\lambda}{2 \kappa_b}}e^{-\frac{\lambda u}{2}}\nonumber \\ &&
+\frac{1}{\sqrt{4\pi\w}}\sum_{n=0}^{\infty} \frac{(-1)^n}{n!}\left(-i\frac{\w}{\kappa_b}\right)^n \frac{ n}{\left( n -\frac{\lambda}{2\kappa_b}\right)} \; e^{-\kappa_b u n} \Big\}\; \theta(r_*) \;.  \eea
Using the identities
\bes \bea
\sum_{n=1}^{\infty} \frac{(i\frac{\w}{\kappa_b} e^{-\kappa_b v})^n }{\left(n-\frac{\lambda}{2 \kappa_b}\right)\; n!}&=&-\frac{2 \kappa_b}{\lambda} \left[-1+e^{i\frac{\w}{\kappa_b} e^{-\kappa_b v}} \right. \nonumber \\
&& \;\;\; \left. + \; \left(-i\frac{\w}{\kappa_b} e^{-\kappa_b v}\right)^{\frac{\lambda}{2 \kappa_b}}\;\gamma\left(1-\frac{\lambda}{2 \kappa_b}, -i\frac{\w}{\kappa_b} e^{-\kappa_b v}\right)\right]\;, \label{inf-sum1}\\
\sum_{n=1}^{\infty} \frac{\left(i\frac{\w}{\kappa_b} e^{-\kappa_b u}\right)^n\; n}{\left(n-\frac{\lambda}{2 \kappa_b} \right)\; n!}&=&-\left(-i\frac{\w}{\kappa_b} e^{-\kappa_b u}\right)^{\frac{\lambda}{2 \kappa_b}}\;\gamma\left(1-\frac{\lambda}{2 \kappa_b}, -i\frac{\w}{\kappa_b} e^{- \kappa_b u}\right)\;, \label{inf-sum2}
\eea \ees
one finds
\bea
p^b_\w &=&\frac{1}{\sqrt{4 \pi \w}}\left[e^{i \frac{\w}{\kappa_b} e^{-\kappa_b u}} - e^{i\frac{\w}{\kappa_b} e^{-\kappa_b v}}+\left(-i\frac{\w}{\kappa_b} e^{-\kappa_b v}\right)^{\frac{\lambda}{2 \kappa_b}} \; \Gamma\left(1-\frac{\lambda}{2 \kappa_b},-i\frac{\w}{\kappa_b} e^{-\kappa_b v}\right)\right]\theta(-r_*)\nonumber \\&&
+\frac{1}{\sqrt{4\pi \w}}\left(-i\frac{\w}{\kappa_b} e^{-\kappa_b u}\right)^{\frac{\lambda}{2 \kappa_b}}\; \Gamma\left(1-\frac{\lambda}{2 \kappa_b},-i\frac{\w}{\kappa_b} e^{-\kappa_b u}\right) \theta(r_*) \;, \label{pb-final-delta}
\eea
where $\gamma(a,z)$ and $\Gamma(a,z)$ are incomplete gamma functions that satisfy the relationship
\be \Gamma(a) =  \gamma(a,z) + \Gamma(a,z) \;. \ee
Note that the infrared divergence in $p^b_\w$ has been removed by scattering effects.

For large enough values of $\w$, one expects scattering effects to be small.  Here the relevant relation is
$\frac{\w}{\kappa} e^{-u/4} \gg 1$.  In this limit
\bea
\Gamma\left(1-\frac{\lambda}{2 \kappa_b},-i\frac{\w}{\kappa_b} e^{-\kappa_b u}\right)=\left(-i\frac{\w}{\kappa_b} e^{-\kappa_b u}\right)^{-\frac{\lambda}{2 \kappa_b}}e^{i\frac{\w}{\kappa_b} e^{- \kappa_b u}}+\mathcal{O}(\w^{-1})\; .
\eea
Substituting this into~\eqref{pb-final-delta}, and letting $u \to v$ where appropriate, gives
\be p^b_\w \approx \frac{1}{\sqrt{4 \pi \w}} e^{i \frac{\w}{\kappa_b} e^{-\kappa_b u}} \ee
for all values of $r_*$.

However, an important caveat is that $\frac{\w}{\kappa} e^{-u/4} \gg 1$ holds for fixed $u$ and large enough $\w$, but it does not hold if $\w$ is fixed and $u$ becomes arbitrarily large.
For a fixed value of $\w$ and a fixed value of $r_*$, one can use the relationship
\be \Gamma(a,z) = \Gamma(a) - \sum_{n=0}^\infty  \frac{(-1)^n z^{a+n}}{n! (a + n)} \;,  \ee
with $a =1-\frac{\lambda}{2 \kappa_b}$ and $z = -i\frac{\w}{\kappa_b} e^{-\kappa_b u}$ or $z = -i\frac{\w}{\kappa_b} e^{-\kappa_b v}$, to show that in the limit $t \to \infty$,  $p^b_\w \to 0$  for all values of $r_*$.  Note that we are assuming here that $a$ is neither zero nor a negative integer.

Recall that, in the case where the potential is zero and there is no scattering, $p^b_\w \to  \frac{1}{\sqrt{4 \pi \w}}$ in the same limit.
So, in this case the scattering has a profound effect on the mode functions.  It was argued in Sec.~\ref{sec:nopot} that the late-time behavior of $p^b_\w$, when there is no scattering, is determined by a pole in the integrand of~\eqref{pb-one-integral} at $\w' = 0$.  When there is scattering due to the potential $V = \lambda \delta(r_*)$, this pole is removed. The resulting integral is of the general form
\be \int_0^\infty d \w' [g(\w',r_*) e^{-i \w' t} + g(-\w', r_*) e^{i \w' t} ] \;, \ee
with $g(\w',r_*)$ having a finite value at $\w' = 0$ and vanishing in the limit $\w' \to \infty$.  For large enough values of $t$,\ it was shown in Sec.~\ref{sec:nopot} that the integrand has no stationary phase point if $V = 0$.  That analysis is valid in this case as well.  In the absence of a stationary phase point, one expects that an integral of this form will vanish in the limit $t \to \infty$ due to the rapid oscillations of the integrand.  This appears to be what is happening.

The same type of late-time behavior has been found for solutions to the 4D wave equation with compact support (and in some cases non compact support) for a massless minimally coupled scalar field in Schwarzschild and other asymptotically flat, static, spherically symmetric spacetimes where scattering occurs due to an effective potential, see e.g.~\cite{angelopoulos,barack} and references contained therein.

\subsection{Massive Scalar Field in Schwarzschild-de Sitter Spacetime}

It was shown in the previous section that when there is scattering of the modes of a massless minimally coupled scalar field in 2D due to a delta function potential, the infrared divergences in both the Boulware modes and the Kruskal modes $p_{\w}^b$ are removed.  This results in the Kruskal modes vanishing at future timelike infinity.  To see whether these results hold for a more realistic potential, we next consider a massive scalar field in 2D. The potential is
\be V  = m^2 f \;,  \label{V-massive}\ee
with $m$ the mass of the field.  In Schwarzschild-de Sitter spacetime the metric function $f$ is
\be f = 1 - \frac{2 M}{r} - H^2 r^2  = -\frac{H^2}{r} (r-r_b) (r - r_c) (r + r_b + r_c) \;, \label{metric-SdS} \ee
with $M$ the mass of the black hole and $H^2 = \frac{1}{3} \Lambda$, where $\Lambda$ is the cosmological constant.  The black hole horizon is at $r = r_b$ and the cosmological horizon is at $r = r_c$.
The quantities $M$ and $H$ can be written in terms of $r_b$ and $r_c$ with the result
\be M = \frac{r_b r_c (r_b+  r_c)}{2(r_b^2 +r_c^2 + r_b r_c)} \;, \qquad H^2 = \frac{1}{r_b^2 + r_c^2 + r_b r_c} \;.\ee
Solving the second relation for $r_c$ gives
\be r_c = -\frac{r_b}{2} + \frac{1}{H}\sqrt{1 - \frac{3 H^2 r_b^2}{4}} \;. \ee
Note that $r_c = 1$ if $r_b = 0$ and that $r_c$ decreases as $r_b$ increases.  They have the same value when $r_b = \frac{1}{\sqrt{3} \; H}$.

As expected $f$ vanishes at the black hole and cosmological horizons $r_b$ and $r_c$.  There is a singularity at $r = 0$ so the spacetime does not extend to negative values of $r$.  However, the function $f$ also vanishes at $r = -r_b-r_c$.  It is useful to compute the surface gravities at each of these zeros of $f$.  If we define them to be positive definite then the results are~\cite{a-t}
  \bea
  \kb &=& {\frac{H^2}{2 r_b} } (r_c -r_b ) (r_c + 2r_b )  \;, \\ \nonumber
    \kc &=& {\frac{H^2}{ 2 r_c} } (r_c -r_b ) (2r_c + r_b )  \;, \\ \nonumber
      \kn &=& {\frac{H^2} {2(r_c +  r_b )} } (2r_c + r_b ) (r_c + 2r_b ) \;.  \label{surfgravs} \eea
The tortoise coordinate can be written in terms of these surface gravities with the result~\cite{a-t}.
\bea\label{rstar}
 r_* (r) &  =& {1\over 2\kb} \log  {| r-r_b | \over r_c - r_b }- {1\over 2\kc} \log  {| r-r_c | \over r_c  -r_b}
 + {1\over 2\kn} \log   {| r+r_c + r_b  | \over r_c + 2r_b } \\ \nonumber
 & - & {r_c \over 4r_b \kb } \log{2r_c + r_b \over r_c +2r_b}  -{r_b r_c \over 2 ( r_c - r_b )} \log{r_b\over r_c} \;.
\eea

Examination of~\eqref{V-massive} and~\eqref{metric-SdS} shows that the potential is zero at the black hole and cosmological horizons, both past and future.  As one goes from the black hole to the cosmological horizon, the potential increases monotonically to a peak and then decreases monotonically to zero.
Because of this, the initial data on the black hole horizon for $h^b_\w$ and $p^b_\w$ and the initial data on the past
cosmological horizon for $h^c_\w$ and $p^c_\w$ is the same for both the massive and massless field.  Thus the solutions are the same as one would find for a massless field with the potential~\eqref{V-massive}.

Due to the complicated form of the potential when expressed in terms of $r_*$, the radial mode functions $\chi^\infty_R$ and $\chi^\infty_L$ have been computed numerically.  They have the forms in~\eqref{chi-R-chi-L-infinity} only on the cosmological horizon where $r_* \to \infty$.
On the black hole horizon, where $r_* \to -\infty$, they take the forms~\eqref{chi-R-L-minfinity}. The scattering coefficients $E_R$, $F_R$, $E_L$, and $F_L$ have been computed numerically for various values of $\w$.

One of the questions we want to address is whether scattering due to this potential removes the infrared divergences in the Boulware modes.  This type of analysis was done in~\cite{rigorous} in terms of the mode functions $\chi^c_\w$ and $\chi^s_\w$, which in the limit $r_* \to \infty$ have the behaviors
\be \chi^c_\w \to \cos(\w r_*) \;, \qquad \chi^s_\w \to \sin(\w r_*) \;, \label{chic-chis} \ee
and in the limit $r_* \to - \infty$ have the form
\bea \chi^c_\w &=& A \cos(\w r_{*}) + B \sin(\w r_{*}) \;, \nonumber \\
     \chi^s_\w &=& C \cos(\w r_{*}) + D \sin(\w r_{*}) \;. \label{ABCD} \eea
It was found in~\cite{rigorous} that for small values of $\w$
\bea A &=& \mathscr{A} + O(\w^2) \;, \nonumber \\
     B &=& \frac{\mathscr{B}}{\w} + O(\w) \;, \nonumber \\
     C &=& \w \, \mathscr{C} + O(\w^3) \;, \nonumber \\
     D &=& \mathscr{D} + O(\w^2) \;. \label{ABCD-small-w} \eea
In the zero frequency limit, there is a divergence in $B$ if $\mathscr{B}$ is nonzero.

To compute the value of $\mathscr{B}$ the relations
\bea E_R &=& \frac{1}{2} [A + D - i(B-C)] \;, \nonumber \\
     E_L &=& \frac{1}{2} [A - D - i (B + C)] \;, \label{EREL-ABCD}  \eea
were used to obtain
\be  B = -{\rm Im} (E_R + E_L)  \ee
numerically for several small values of $\w$.  Then the quantity $\w B$ was fitted to a power series in $\w$, and it was found in the limit $\w \to 0$ that for $H m = 1$,  $\mathscr{B} \approx -1.45$.

We next show that if $\mathscr{B} \ne 0$, then the infrared divergences in the Boulware modes $h^b_\w$ and $h^{(\mathscr{I}^{-} \, c)}_\w$ are removed.
Using~\eqref{ABCD-small-w} and~\eqref{EREL-ABCD}, one finds in the low frequency limit that
\bea \frac{1}{E_R} &\to& \frac{2 i \w}{\mathscr{B}} \;, \nonumber \\
          \frac{E_L}{E_R} &\to& 1 - 2 i \w \frac{\mathscr{D}}{\mathscr{B}} \;. \label{div-removal} \eea
It was shown in~\cite{rigorous} that in this limit
\bea \chi^\infty_R &\to& \chi^{(1)}_0 + i \w \chi^{(2)}_0 \;, \nonumber \\
     \chi^\infty_L &\to& \chi^{(1)}_0 - i \w \chi^{(2)}_0 \;. \label{chi01-chi02-1} \eea
Here $\chi^{(1)}_0$ is the solution to the mode equation~\eqref{mode-eq-chi} with $\w = 0$ that is approached by $\chi^c_\w$ in the limit $\w \to 0$, and
$\chi^{(2)}_0$ is the solution to the same equation that is approached by $\frac{\chi^s_\w}{\w}$ in the limit $\w \to 0$.
Substitution of~\eqref{div-removal} and~\eqref{chi01-chi02-1} into~\eqref{chi-b-I-c-scatt} and then substituting the result into~\eqref{modes-sep-var} gives
\bea h^b_\w  &=& \frac{i}{\mathscr{B}} \sqrt{\frac{\w}{\pi}} \, \chi^{(1)}_0 + O(\w^{3/2}) \;, \nonumber \\
     h^c_\w  &=& i\sqrt{\frac{\w}{\pi}} \left[\frac{\mathscr{D}}{\mathscr{B}} \, \chi^{(1)}_0 - \chi^{(2)}_0 \right] + O(\w^{3/2}) \;.
     \label{hb-hc-small-w} \eea

\begin{figure}[h]
\centering
\includegraphics [trim=0cm 0cm 0cm 0cm,clip=true,totalheight=0.20\textheight]{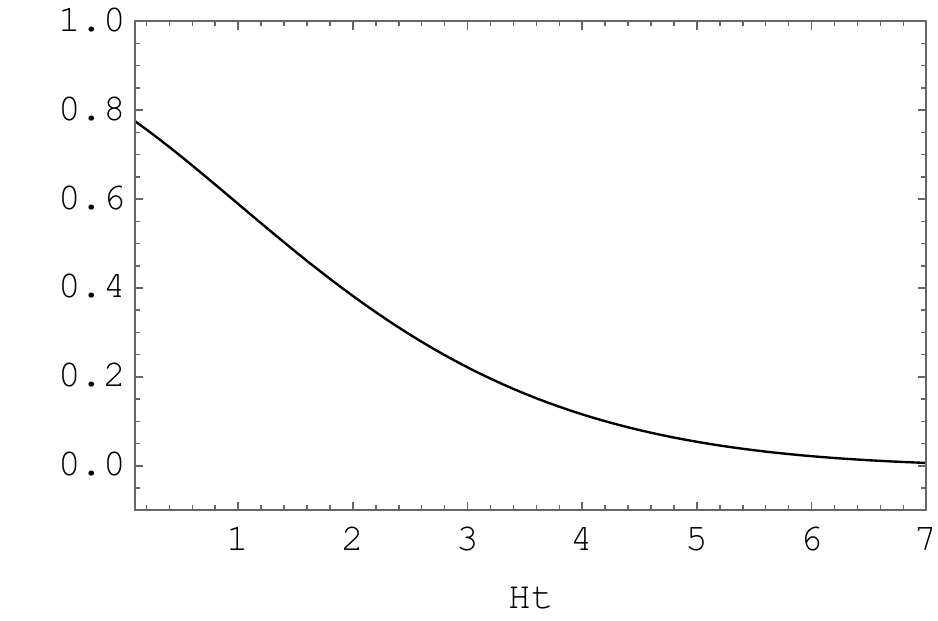}
\includegraphics [trim=0cm 0cm 0cm 0cm,clip=true,totalheight=0.20\textheight]{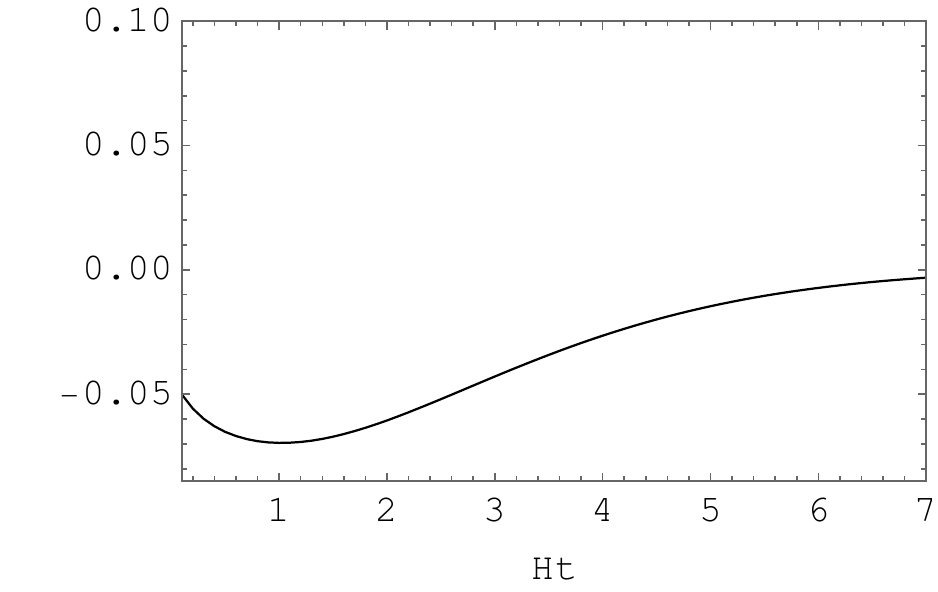}
\caption{The plot on the left shows the real part of $\sqrt{H} p^b_\w$ for $ H r_b = \frac{1}{10}$, $\w = \frac{H}{10}$, $H r = \frac{3}{10}$ and various times $H t$. The plot on the right shows the imaginary part of $\sqrt{H} p^b_\w$.  Both the real and imaginary parts of $p^b_\w$ approach zero at late times.  }
\label{fig:pb}
\end{figure}

We next consider the late-time behavior of $p^b_\w$.  That of $p^c_\w$ is similar.
To compute $p^b_\w$, the results of the numerical computations of $\chi^\infty_R$ and $\chi^\infty_L$ were substituted into~\eqref{pbeq} and
the integral was numerically computed at various times using~\eqref{alpha-beta-b}.
The results show that $p^b_\w$ approaches zero in the large $t$ limit for fixed $r$.  This is illustrated in Fig.~\ref{fig:pb},
for the particular case $m H = 1$ and $ \w = 0.1 H$.  It is clear that both the real and imaginary parts of $p^b_\w$ approach zero in the late-time limit.  This behavior is the same type of behavior as occurs for the delta function potential.

\section{Two-Point Function}

 In~\cite{a-t} it was shown that for a massless minimally coupled scalar field in a spacetime with a region in which the 2D metric is static and there is either a black hole horizon, a cosmological horizon, or both, linear growth occurs in $G^{(1)}(x,x')$ in terms of a time coordinate $T$ that is well-behaved on the future horizon(s) if the field is in the Unruh vacuum state and the points are split in the radial direction.  In this section we first review that work and then show that the result implies that, in this case, the two formulations of the Unruh state given in~\cite{unruh:1976} (see also~\cite{fulling:1977}) give inequivalent answers for $G^{(1)}$ in an asymptotically flat black hole spacetime  with a 2D metric for a massless minimally coupled scalar field.  We tie this to the existence of an infrared divergence in the Boulware mode functions that affects the order in which three integrals used to evaluate $G^{(1)}$ are computed.  We next investigate the behavior of $G^{(1)}$ in a black hole spacetime  with a 2D metric when there is scattering due to a delta function potential.  We show that if the points are separated in the radial direction then there is no time dependence, and we show numerically that the two formulations of the Unruh state in~\cite{unruh:1976} are equivalent.

The symmetric two-point function for a scalar field is
\be G^{(1)}(x,x') = \la 0| \phi(x) \phi(x') + \phi(x') \phi(x)| 0\ra \;. \label{G1-def} \ee
For the Unruh state, the field can be expanded in terms of the complete set of modes $p^b_\w$ and $h^{\mathscr{I}^{-}}_\w$, with the result
\be \phi = \int_0^\infty d\w \left[ a^b_\w p^b_\w + a^{b \dagger}_\w p^{b \, *}_\w + a^{\mathscr{I}^{-}}_\w h^{\mathscr{I}^{-}}_\w +
   a^{\mathscr{I}^{-} \dagger}_\w h^{\mathscr{I}^{-} *}_\w \right] \;. \label{phi-first} \ee
Substituting this into~\eqref{G1-def} gives
\bea G^{(1)}(x,x') &=& \int_0^\infty d \w [p^b_\w(x) p^{b *}_\w(x') + p^b_\w(x')  p^{b *}_\w(x) \nonumber \\
  & &  + h^{\mathscr{I}^{-}}_\w(x)
 h^{\mathscr{I}^{-}*}_\w(x') +  h^{\mathscr{I}^{-}}_\w(x')
 h^{\mathscr{I}^{-}*}_\w(x)]\;. \label{G-first}\eea
A second formulation~\cite{unruh:1976} involves expanding the field in terms of the modes $h^b_\w$ and $h^{\mathscr{I}^{-}\, *}_\w $ but with a normalization that leads to a thermal distribution
\bea \phi &=& \int_{-\infty}^\infty d\w \left[ (a^b_\w)_{th} (h^b_\w)_{th} + (a^{b \, \dagger}_\w)_{th} (h^{b \, *}_\w)_{th} \right]
   + \int_0^\infty d\w \left[  a_\w^{\mathscr{I}^{-}} h^{\mathscr{I}^{-}}_\w +
   a_\w^{\mathscr{I}^{-}\, \dagger} h^{\mathscr{I}^{-}\, *}_\w \right] \;, \label{phi-second} \eea
with
\be (h^b_\w)_{th} =  \frac{e^{\frac{\pi\w}{2 \kappa_b}}}{[2\sinh{(\pi \w/\kappa_b)}]^{\frac{1}{2}}} \; h^b_\w \;. \label{hbth}  \ee
Substitution into~\eqref{G1-def} gives
\bea G^{(1)}(x,x') &=& \int_{-\infty}^\infty d \w \; \frac{e^{\frac{\pi \w}{\kappa_b}}}{2 \sinh \left( \frac{\pi \w}{\kappa_b} \right)}  \left[h^b_\w(x) h^{b \, *}_\w(x') +  h^b_\w(x') h^{b \, *}_\w(x) \right]  \nonumber \\
 &&  + \int_0^\infty [ h^{\mathscr{I}^{-}}_\w(x)
 h^{\mathscr{I}^{-}*}_\w(x') +  h^{\mathscr{I}^{-}}_\w(x')
 h^{\mathscr{I}^{-}*}_\w(x) ] \;. \label{G-second}\eea

\subsection{Connection between the two formulations of the Unruh state}

One way to connect the two formulations of the Unruh state is to start with~\eqref{G-first} and use the Bogolubov transformation~\eqref{pbeq} with~\eqref{alpha-beta-b}.  The result for the part that depends on the Kruskal modes $p^b_\w$ is
\bea G^{(1)}_K(x,x') &=& \int_0^\infty d \w \int_0^\infty d \w_1 \int_0^\infty d \w_2 \left\{ \alpha^b_{\w \w_1}  \alpha^{b \, *}_{\w \w_2}
[h^b_{\w_1}(x) h^{b \, *}_{\w_2}(x') + h^b_{\w_1}(x') h^{b \, *}_{\w_2}(x)] \right. \nonumber \\
& & \left.  + \beta^b_{\w \w_1}  \beta^{b \, *}_{\w \w_2}
[h^{b \, *}_{\w_1}(x) h^{b}_{\w_2}(x') + h^{b \, *}_{\w_1}(x') h^{b}_{\w_2}(x)] \right. \nonumber \\
  && \left.  + 2 {\rm Re} \left( \alpha^b_{\w \w_1}  \beta^{b \, *}_{\w \w_2} [h^b_{\w_1}(x) h^{b}_{\w_2}(x') + h^b_{\w_1}(x') h^{b}_{\w_2}(x)] \right) \right\} \;. \label{GK1} \eea
If the order of integration is changed so that the integral over $\w$ is done first, then
\bea G^{(1)}_K(x,x') &=& \int_0^\infty d \w_1 \int_0^\infty d \w_2 \left\{
[h^b_{\w_1}(x) h^{b \, *}_{\w_2}(x') + h^b_{\w_1}(x') h^{b \, *}_{\w_2}(x)] \; I_1 \right. \nonumber \\
&& \left.  +
[h^{b \, *}_{\w_1}(x) h^{b}_{\w_2}(x') + h^{b \, *}_{\w_1}(x') h^{b}_{\w_2}(x)]  \;  I_2  \right. \nonumber \\
  && \left.  + 2 {\rm Re} \left(  [h^b_{\w_1}(x) h^{b}_{\w_2}(x') + h^b_{\w_1}(x') h^{b}_{\w_2}(x)] \;I_3 \right) \right\} \;, \label{GK2} \eea
with
\bes \bea I1 &=&  \int_0^\infty d \w \alpha^b_{\w \w_1}  \alpha^{b \, *}_{\w \w_2} \;,\label{I1} \\
    I2 &=&  \int_0^\infty d \w \beta^b_{\w \w_1}  \beta^{b \, *}_{\w \w_2} \;,\label{I2} \\
     I3 &=&  \int_0^\infty d \w  \alpha^b_{\w \w_1}  \beta^{b \, *}_{\w \w_2}  \;. \label{I3} \eea \label{I1-I3} \ees
Substituting~\eqref{alpha-beta-b} into~\eqref{I1} gives \bea I_1 &=& \frac{\sqrt{\w_1 \w_2}}{4 \pi^2 \kappa_b^2} e^{\frac{\pi}{2 \kappa}(\w_1+\w_2)} \kappa^{-\frac{I}{\kappa_b}(\w_1-\w_2)} \Gamma\left(\frac{-i \w_1}{\kappa_b} \right) \Gamma\left(\frac{i \w_2}{\kappa_b} \right) \int_0^\infty \frac{d \w}{\w} \w^{\frac{i}{\kappa_b}(\w_1-\w_2)} \;. \label{I1-2} \eea
Changing the integration variable to $z = \log \w$, one easily finds that the integral is equal to $2 \pi \kappa_b \delta(\w_1 - \w_2)$.  One finds a similar result for $I_2$.  However, for $I_3$ the corresponding integral is proportional to $\delta(\w_1 + \w_2)$, which is zero when integrated over because the integrals are over nonnegative values of $\w_1$ and $\w_2$.  The final result is that $G^{(1)}(x,x')$ is
the same as was found for the second formulation of the Unruh state, which is displayed in~\eqref{G-second}.

\subsection{No Scattering Case}

There is no scattering for the $m = \xi = 0$ scalar field in 2D
, so, as discussed in Sec.~\ref{sec:modes}, we have analytic expressions for the mode functions throughout the spacetime.  For an asymptotically flat black hole spacetime the symmetric two-point function in the first formulation~\eqref{G-first} can be computed analytically with the result~\cite{a-t}
\bea G^{(1)}(x,x')  &=&  - \frac{1}{2 \pi} \left\{ {\rm ci}(\w_0|U_b-U_b'|) +  {\rm ci}(\w_0|v-v'|) \right\} \nonumber \\
    & = & - \frac{1}{2 \pi}\left\{ \log( \omega_0 |U_b-U_b'|) + \log(\omega_0|v-v'|)  + 2 \gamma_E  \right\} \;.
\eea
Here $\gamma_E$ is Euler's constant and $\w_0$ is an infrared cutoff that is necessary because of the infrared divergence in the integrand
of~\eqref{G-first} in this case.  Note that the second expression works so long as
$\w_0 |U_b - U_b'| \ll 1$ and $\w_0|v-v'| \ll 1$.
In the region outside the black hole horizon
\be U_b -U_b' = - \frac{1}{\kappa_b} (e^{-\kappa_b u} - e^{-\kappa_b u'}) = -\frac{e^{-\kappa_b u}}{\kappa_b} (1 - e^{\kappa_b (u - u')}) \;. \ee
One finds that
\bea G^{(1)}(x,x')  &=& \frac{1}{2 \pi} \left\{ - 2 \gamma_E + \kappa_b u  - \log \left(\frac{\w_0}{\kappa_b} |1- e^{\kappa_b (u-u')}| \right) - \log (\omega_0|v-v'|) \right\} \;.
\eea

There is clearly a linear growth in the null coordinate $u$ for fixed $u - u'$ that is proportional to the surface gravity of the black hole.  This means that for fixed $r' \ne r$ and for $t' = t$, there is linear growth in time that is proportional to the surface gravity of the black hole.  If one defines a different time coordinate through the relation $T = t + q(r)$ for some function $q(r)$ then there is linear growth in $T$
if $T' = T$ and $r' \ne r$.  Similarly, for fixed $t' \ne t$ (or $T' \ne T$) and for $r'=r$, there is linear growth as the radial coordinate approaches the horizon and a linear growth in the negative value of $G^{(1)}(x,x')$ as the radial coordinate approaches infinity.

For the second formulation~\eqref{G-second}, we obtain
\bea G^{(1)}(x,x') = -\frac{1}{2 \pi} {\rm ci}(\w_0|v-v'|) + \int_{\w_0}^\infty \frac{d \w}{\w} {\rm coth} \left(\frac{\pi \w}{\w} \right) \cos [\w (u-u')] \;. \eea
In this case there is clearly no growth in $u$ when $u-u'$ is held fixed and thus no linear growth in $t$ when $t' = t$, and $r$ and $r'$ are fixed.  Therefore the two formulations are not equivalent.
The likely reason is that there are infrared divergences in the integrands of the
integrals over $\w_1$ and $\w_2$ in~\eqref{GK1}.  This can be seen by setting $\delta = 0$ in~\eqref{alpha-beta-b}.  These infrared divergences probably
prevent one from interchanging the order of integration as was done to obtain~\eqref{GK2}.

\subsection{Delta Function Potential}

For the delta function potential, scattering occurs and the infrared divergences in the mode functions $p^b_\w$, $h^b_\w$, and $h^{\mathscr{I^{-}}}_{\w}$ are removed.  Thus one would expect that it is possible to interchange the order of the integrals in~\eqref{GK1}.
We verify this conjecture by explicit calculation here.  Our treatment is valid for any static, asymptotically flat  black hole spacetime  with a 2D metric.

The radial part of the mode equation was solved exactly in~\eqref{h-b-delta} and~\eqref{h-I-c-delta} for the delta function potential. For the first formulation, we can use this result in~\eqref{GK1} for $r_* > 0$ to obtain
\bea
G^{(1)}_K(x,x') &=& 2{\rm Re} \left\{ \int_0^{\infty}d\w\; p_{\w}^b (x) p_{\w}^{b\;*}(x')\right\}  \nonumber \\
&=& \frac{1}{2\pi}{\rm Re} \left\{ \int_0^{\infty}\frac{d\w}{\w}(-i\frac{\w}{\kappa_b} e^{-\kappa_b u})^{\frac{\lambda}{2 \kappa_b}}\Gamma\left(1-\frac{\lambda}{2 \kappa_b}, -i\frac{\w}{\kappa_b}e^{-\kappa_b u}\right) \right. \nonumber  \\  && \left.
 \times(i\frac{\w}{\kappa_b} e^{-\kappa_b u'})^{\frac{\lambda}{2 \kappa_b}}\Gamma\left(1-\frac{\lambda}{2 \kappa_b}, i\frac{\w}{\kappa_b}e^{-\kappa_b u'}\right)\right\} \;. \label{G-first-formulation}
\eea
Changing the variable of integration such that $z=\frac{\w}{\kappa_b} e^{-\kappa_b u}$,~\eqref{G-first} gives
\bea
G^{(1)}_K(x,x') &=&2{\rm Re}\left\{\int_0^\infty d\w\; p_{\w}^b (x) p_{\w}^{b\;*}(x')\right\}\nonumber \\
&=&\frac{1}{2\pi} {\rm Re} \left\{\int_0^{\infty}\frac{dz}{z} z^{\frac{\lambda}{\kappa_b}}e^{\frac{\lambda}{2} \Delta u}\Gamma\left(1-\frac{\lambda}{2 \kappa_b},-iz\right)\; \Gamma\left(1-\frac{\lambda}{2 \kappa_b}, i z e^{\kappa_b \Delta u}\right) \right\} \,, \label{G-first-scaled}
\eea
where $\Delta u = u - u'$.

\begin{figure}[h]
\centering
\includegraphics [trim=0cm 0cm 0cm 0cm,clip=true,totalheight=0.20\textheight]{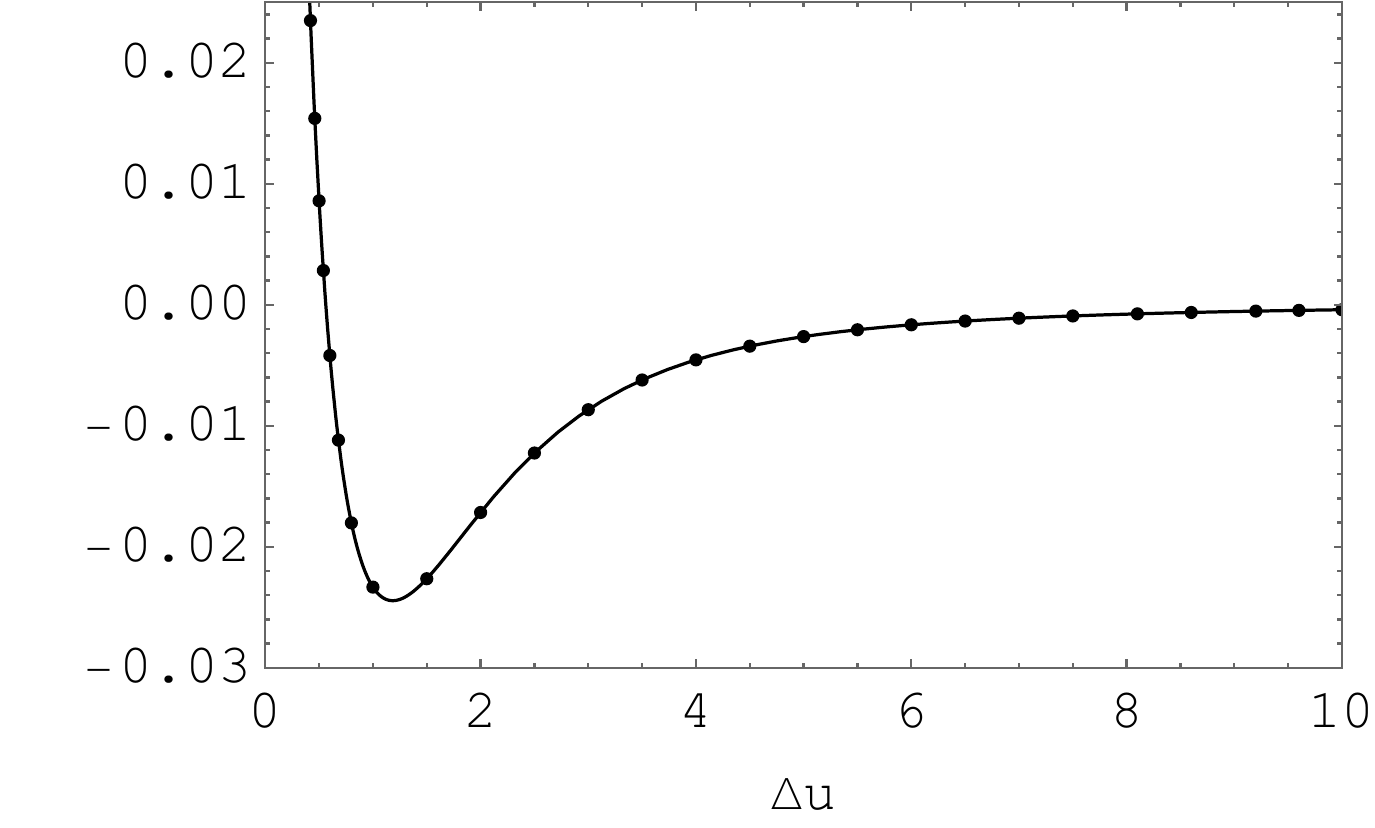}
\hspace{1cm}
\caption{Plotted is the symmetric two-point function $G_K^{(1)}(x,x')$ for the delta function potential as a function of $\Delta u$ when $\lambda = \pi$, and both $r_* > 0$ and $r^{'}_* > 0$.  The dots result from numerical evaluations using the first formulation for the Unruh state~\eqref{G-first-scaled} and the solid curve is for the second formulation~\eqref{GK-method-2}.  It is clear that the two formulations give the same result for the symmetric two-point function in this case.}
\label{fig:comparison}
\end{figure}

Thus, for $r_*>0$, it is clear that there is no time dependence if $t' = t$ and, in general, that the two-point function is simply a function of $\Delta u$ and $\Delta v = v - v'$.  It is interesting to note that, as shown in Fig.~\ref{fig:comparison}, there is a negative correlation peak in $G^{(1)}_K(x,x')$ at a particular value of $\Delta u$.  For $r_* < 0$, it is not obvious from the form of the integrand for $G^{(1)}_K(x,x')$ what happens when $t' = t$.  However, numerical computations at different times when $t' = t$ and the radial points are split show that there is no time dependence in this case.

For the second formulation, one sees from~\eqref{G-second} that for $r_* > 0$
\bea
G^{(1)}_K(x,x')&=& \int_{-\infty}^\infty d \w \;\frac{ e^{\frac{\pi \w}{\kappa_b}}}{2 \sinh \left(\frac{\pi \w}{2 \kappa_b} \right)}  \left[h^b_\w(x) h^{b \, *}_\w(x') +  h^b_\w(x') h^{b \, *}_\w(x) \right] \nonumber \\
              &=& \frac{1}{4 \pi} {\rm Re} \left\{ \int_{-\infty}^\infty d \w \frac{\w }{\sinh\left(\frac{\pi \w}{\kappa_b} \right) } \frac{e^{\frac{\pi \w}{\kappa_b}} \; e^{i \w \Delta u}}{\left(\w^2 + \frac{\lambda^2}{4} \right)}  \right\} \;. \label{G-second-K}
\eea
The integral can be evaluated using standard complex integration techniques.  For either sign of $\Delta u$ and for $\lambda \ne 2 n \kappa_b$ with $n$ an integer, we find
\bea
G^{(1)}_K(x,x')&=& \frac{1}{2 \pi}  \sum_{n=1}^\infty \left[ \frac{n}{\left(n^2- \frac{\lambda^2}{4 \kappa_b^2} \right)} \; e^{- \kappa_b |\Delta u| n} \right] + \frac{1}{4} {\rm Re} \left\{ \frac{e^{i \frac{\pi \lambda}{2 \kappa_b}}}{\sin \left(\frac{\pi \lambda}{2 \kappa_b} \right)}   \; e^{-\frac{\lambda}{2} |\Delta u|}  \right\} \nonumber \\
& = & \frac{ \kappa_b \; e^{-\kappa_b |\Delta u|}}{ 2 \pi \lambda} \Bigg\{\frac{1}{ \left(1- \frac{\lambda}{2 \kappa_b} \right)}\; _2 F_1\left(2,1-\frac{\lambda}{2 \kappa_b};2-\frac{\lambda}{2 \kappa_b};e^{- \kappa_b |\Delta u|}\right)\nonumber \\&&
- \frac{1}{ \left( 1 + \frac{\lambda}{2 \kappa_b}\right)}\; _2F_1\left(2,1+\frac{\lambda}{2 \kappa_b};2+\frac{\lambda}{2 \kappa_b};e^{-\kappa_b |\Delta u|}\right)\Bigg\} \nonumber \\
& &  + \frac{1}{4} \cot \left(\frac{\pi \lambda}{2 \kappa_b} \right) \; e^{\frac{-\lambda}{2} |\Delta u|} \;. \label{GK-method-2}
\eea
 Here $_2 F_1$ is the hypergeometric function which has the power series representation
\bea
_2 F_1(a,b,c;z)=\sum_{n=0}^{\infty} \frac{(a)_n (b)_n}{(c)_n}\frac{z^n}{n!} \; \label{hyper-Geom-Series}
\eea
with $(x)_n\equiv\frac{\Gamma(x+n)}{\Gamma(x)}$ the rising factorial.  The series in ~\eqref{hyper-Geom-Series} converges for $|z|<1$.  It is straight-forward to verify the results in~\eqref{GK-method-2} if one starts with the answer and works backward to the original sum.

Numerical computations of ~\eqref{G-first-scaled} and numerical evaluations of~\eqref{GK-method-2} when both $r_* > 0$ and $r^{'}_* > 0$  indicate that these two expressions are the same for the
delta function potential.  Our results are shown in Fig.~\ref{fig:comparison} for the case $\lambda = \pi$.

As shown in the previous section, the two formulations of the Unruh state result in very different symmetric two-point functions in 2D when there is no scattering.  However, for $r_*>0$ they give the same two-point function when there is scattering due to a delta function potential; we expect to get the same result for $r_* < 0$. The difference appears to be that the removal of the infrared divergences in~\eqref{GK1} by scattering effects allows for the interchange of the order of integration in~\eqref{GK2} and the resulting equivalence of the two formulations.

It is interesting to note that the dominant contribution to $G^{(1)}_K(x,x')$ in the first formulation~\eqref{G-first-scaled} comes from intermediate values of the variable $z$, which implies that for very large values of $u$ with fixed $\Delta u$ (for example, $t'=t$, fixed values of $r$ and $r'$, and very large values of $t$), they come from extremely high frequencies $\w$.  Conversely, in the second formulation~\eqref{G-second-K}, they come from intermediate values of $\w$ at all times.   This is reminiscent of the transplanckian issue for Hawking radiation.  There, the {\it in} modes that contribute significantly to the radiation at distances far from the black hole at late times have frequencies that are much higher than the Planck scale, while the frequencies of the particles that would be detected are much lower than the Planck scale.  Using a Bogolubov transformation between the {\it in} and {\it out} vacuum states, one can write the {\it in} modes in terms of a wave packet of the {\it out} modes.  At late times (how late depends on the frequency of the {\it in} mode), a high frequency {\it in} mode gets its main contributions from {\it out} modes that have frequencies well below the Planck scale.

\section{Summary and Conclusions}

We have investigated the properties of the modes and the symmetric two-point function for scalar fields in the Unruh state in black hole spacetimes with 2D metrics with and without cosmological horizons.  Our focus has been outside of the black hole event horizon and inside the cosmological horizon, if one exists.
For simplicity, we have presented the results of specific calculations for the Kruskal modes associated with the past black hole horizon, $p^b_\w$.  If a cosmological horizon is present, the calculations for the Kruskal modes associated with the past cosmological horizon $p^c_\w$ are similar.
 We have considered three different cases:  a massless minimally coupled scalar field in any static black hole spacetime  with a 2D metric, a massless minimally coupled scalar field with a potential in the mode equation of the form $V = \lambda \delta(r_*)$ in asymptotically flat static black hole spacetimes  with 2D metrics, and a massive minimally coupled scalar field in SdS spacetime with a 2D metric.

There are infrared divergences associated with normalizing the Boulware modes and the Kruskal modes.  In asymptotically flat static black hole spacetimes with 2D metrics, we have found that scattering effects due to a delta function potential remove these divergences.  We have also found that scattering that occurs for a massive minimally coupled scalar field in SdS spacetime with a 2D metric removes the infrared divergences in the Boulware modes and the Kruskal modes that are associated with the past black hole horizon.

The existence or nonexistence of infrared diverges for the Boulware modes has a strong effect on the late-time behaviors of the Kruskal modes.  In particular, if there is no scattering so that the infrared divergences are present, then for a fixed value of the radial coordinate $r$, $p^b_\w \to (4 \pi \w)^{1/2}$ at late times. For a massless scalar field with a delta function potential in any eternal black hole spacetime with a 2D metric, and for a massive scalar field in SdS spacetime with a 2D metric, the Boulware mode functions $h_{\w}^b$ are infrared finite and we find that $p^b_{\w} \to 0$ for fixed $r$ in the limit $t \to \infty$.  This also occurs for any other time coordinate $T = t + q(r)$ for any function $q(r)$.

The fact that the Kruskal modes approach zero in the late-time limit for fixed $r$ when the infrared divergences in the Boulware modes are removed is not too surprising. It is well known in Schwarzschild and other 4D asymptotically flat static spherically symmetric spacetimes, that solutions to the wave equation for a classical minimally coupled massless scalar field for data with compact support (and in some cases with non compact support) vanish in the late-time limit for fixed $r$, see e.g.~\cite{angelopoulos,barack} and references therein.

We have also investigated the behavior of the two-point function for the Unruh state.  If one uses the first formulation of the Unruh state~\eqref{phi-first}, the resulting integral in~\eqref{G-first} looks significantly different than that found in~\eqref{G-second}  for the second formulation~\eqref{phi-second}.  However,
if one uses the Bogolubov transformation~\eqref{pbeq}, then the first formulation does give the same answer as the second one if the order of the integrals in~\eqref{GK1} is changed so that the integral over $\w$ is done first.

In~\cite{a-t}, it was shown that for a massless minimally coupled scalar
field in a spacetime with a 2D metric containing a static patch and one or two horizons, the symmetric two-point function grows linearly in time at late times when the time points are equal and the space points are separated and held fixed.  We have shown that this occurs only for the first formulation of the Unruh state~\eqref{phi-first}, where the $p^b_\w$ modes are directly used to compute the two-point function.  For the second formulation~\eqref{phi-second}, there is no such linear growth in time.  The likely reason for the difference is that when there is no scattering, there are infrared divergences in the integrals over $\w_1$ and $\w_2$ in~\eqref{GK1}.  These probably make it invalid to interchange the order of integration, which is the process that makes the first and second formulations of the Unruh state give the same answer for the two-point function. It is the first formulation that is correct in this case.

For the second formulation of the Unruh state,~\eqref{G-second}, we were able to compute the two-point function analytically for $r_*>0$.  We show in Fig.~\ref{fig:comparison} that the values of the two-point function in the two formulations are the same for $r_*>0$.  Note that in the second formulation, the dominant contribution to the integral comes from intermediate values of the frequency, not transplanckian ones.

For the massless scalar field in black hole spacetimes with 2D metrics, the stress-energy tensor does not grow in time when the field is in the Unruh state even though the two-point function does.  Mathematically, the reason appears to be that the stress-energy tensor is obtained by taking various pairs of derivatives of the two-point function, one at the point $x$ and one at the point $x'$ when the points are separated in both the space and time directions.  This keeps any terms that are linear in either the space or time coordinates from contributing to the stress-energy tensor.
Another way to understand this is to apply the derivatives to~\eqref{GK1}.  If no scattering occurs, then these derivatives bring down factors of $\omega_1$ and $\omega_2$ removing the infrared divergences in those integrands.  Then interchanging the order of integration is allowed.

One last point is that when the order of integration can be interchanged, the equivalence between the two formulations still relies upon the infrared divergence of the Kruskal modes on the past horizon.  In~\eqref{I1-2}, the variable transformation $z = \log \w$ puts the integral in the form of the integral for a delta function.  If there was no infrared divergence, then an extra factor of
$e^z$ would occur in the integrand when the change of variables was made.  This would change the value of the integral and that would mean that the two formulations for the Unruh state are not equivalent.

One can ask: which of the effects discussed in this paper are likely to survive in more realistic 4D calculations?  For the massless minimally coupled scalar field, there is always an effective potential in the radial mode equation in a black hole spacetime with a static patch.  Thus scattering effects occur.  For Schwarzschild and Reissner-Nordstrom spacetimes, scattering effects are known to remove the infrared divergences in the Boulware mode functions.  One then expects the Kruskal modes to approach zero at late times and that there should be no linear growth in time of the symmetric two-point function for the Unruh state.  However, for the spherically symmetric modes  (the $\ell = m = 0$ spherical harmonic, $Y_{\ell m}(\theta,\phi)$)  in 4D SdS spacetime, it is known that the infrared divergences in the Boulware mode functions are not removed (although they are for higher values of $\ell$).  Thus one expects the corresponding Kruskal modes to approach nonzero constants at late times and that there should be linear growth in time of this two-point function.  Work is in progress to verify these conjectures.

\newpage

\noindent {\bf Data availability statement}

\vspace{0.3cm}

\noindent The data that supports the findings of this study are openly available at the following URL/DOI: \newline
\noindent http://users.wfu.edu/anderson/research/downloads/index.htm

\acknowledgments

We would like to thank Gregory Cook, Alessandro Fabbri, John Gemmer, and Jennie Traschen for helpful conversations.  This work was supported in part by the National Science Foundation under Grant No. PHY-1912584 to Wake Forest University.

\end{document}